\documentclass[a4paper,10pt]{article}
\usepackage{amssymb,epsfig}

\begin{document}
\topmargin=-1.5cm
\oddsidemargin=-0.5cm
\evensidemargin=-0.5cm

\begin{flushright}
{\small KAIST-TH 2003/16}
\end{flushright}
\begin{center}
{\Large \bf Dynamical symmetry breaking on a brane with bulk gauge theory}

\vspace*{0.5cm}

{\large Hiroyuki~Abe\footnote{E-mail: abe@muon.kaist.ac.kr}} \\

\vspace*{0.2cm}

{\normalsize \it Department of Physics, Korea Advanced Institute 
of Science and Technology, Daejeon 305-701, Korea}

\vspace*{0.2cm}

Talk given at Summer Institute '03, Yamanashi, Japan (12-19 August, 2003)

\begin{abstract}
\noindent
We analyze a structure of dynamical chiral symmetry breaking 
in orbifold gauge theories with matter fields (fermions) on the fixed point. 
We find that the boundary chiral phase structure of QED and QCD on 
the orbifold is quite nontrivial depending on the bulk constitution, 
and we claim that particular attention should be given to the dynamically 
generated masses in various kinds of phenomenological orbifold models. 
\end{abstract}

\end{center}

\vspace*{0.2cm}

%\section{Introduction}
\noindent 
It is revealed recently that the orbifold field theory 
can provide, e.g., the weak and Planck hierarchy~\cite{Randall:1999ee},  
a fermion mass hierarchy (via localization~\cite{Arkani-Hamed:1999dc}), 
a gauge symmetry and a supersymmetry (SUSY) breaking 
\cite{Antoniadis:1990ew,Antoniadis:1993jp} 
(by the Wilson line~\cite{vonGersdorff:2001ak}), 
SUSY breaking mediation mechanisms~(e.g., \cite{Choi:2003di}), 
and the proton stability~(e.g., \cite{Kawamura:2000ev}), 
based on perturbative (or tree) analyses. 
However an effective higher-dimensional theory such as the 
orbifold model means $M_c < \Lambda$ by definition, i.e., the 
compactification scale should be less than the cut-off scale of the theory. 
This implies an existence of Kaluza-Klein particles below $\Lambda$ 
and we have a question about their effect on a nonperturbative dynamics 
of the theory. 

One simple example for such a nonperturbative dynamics is 
a chiral symmetry breaking structure in a $SU(N)$ Yang-Milles 
and matter (fermion) theory on the orbifold. 
Assuming only a kink-type mass $\epsilon(y)M_{\rm kink}$, 
we have a chiral symmetry for the fermion zero mode. 
The fermion zero mode couples to KK gauge bosons  
as well as a gauge boson zero mode. So the chiral phase structure 
may be different from the usual 4D case 
(e.g., a four-fermion approximation, Ref.~\cite{Dobrescu:1998dg}). 
The simplest case for the analysis is $|M_{\rm kink}| \to \Lambda$ 
that means all the KK fermions decouple and the fermion becomes 
a brane field effectively. Also a lot of orbifold models introduces 
intrinsic brane fields (fermions) which couples to a bulk gauge boson. 
Therefore we analyze a dynamical chiral symmetry breaking (DSB) on 
a boundary with a bulk gauge theory. 

%\section{Gauge theory with boundary matter}
We consider QCD (QED) on $M_4 \times S^1/Z_2$ 
with quarks (electrons) on the fixed point. 
For generality, the bulk geometry is assumed as, 
$ds^2=G_{MN}dx^Mdx^N=e^{-2k|y|}\eta_{\mu\nu}dx^\mu dx^\nu - dy^2$, 
where $k$ is a AdS curvature scale. We use $R$, $\Lambda$ 
and $\Lambda_{\rm 5D}$ as a radius of the fifth dimension, 
a 4D (brane) effective cut-off scale and 
a 5D cut-off scale respectively. 
The 4D effective Lagrangian is given by 
\begin{eqnarray}
{\cal L} 
&=& \frac{1}{2} \sum_{n=0}^{N_{\rm KK}} A_\mu^{(n)} \left[ 
    \eta^{\mu \nu} \left( \partial^2 + M_n^2 \right)
   -(1-\xi) \partial_\mu \partial_\nu \right] A_\nu^{(n)} 
%\nonumber \\ && 
   +\bar{\psi} \left( i\partial_\mu + g A_\mu^{(0)} \right) 
    \gamma^\mu \psi
   + g \sum_{n =1}^{N_{\rm KK}} 
       \frac{\chi_n(y^\ast)}{\chi_0(y^\ast)}  
    \bar{\psi} A_\mu^{(n)} \gamma^\mu \psi, \nonumber 
\end{eqnarray}
where $y^\ast = 0, \pi R$ represents orbifold fixed points. 
We neglected gluon self interactions which will be 
incorporated latter via a one-loop running coupling. 
The $\chi_n(y)$ and $M_n$ is the $n$-th Kaluza-Klein (KK) mode function 
and the mass eigenvalue of the gauge field, respectively. 
The $\chi_n(y)$ takes a value of $\chi_n(y^\ast)=\sqrt{2}$ in a flat 
geometry, and less (more) than this value at $y^\ast =0$ ($\pi R$) 
in a warped geometry, depending on the AdS curvature $k$. 
In the following analysis we take $kR=11.35$ as a typical example. 
The $\xi$ is a gauge fixing parameter which will be fixed in such 
a way that the chiral Ward identity is approximately held 
in the following numerical analysis. 

%\section{Dynamical chiral symmetry breaking on a boundary}
To analyze DSB in our system, we solve a Schwinger-Dyson (SD) 
equation for a fermion propagator on the brane, 
\begin{eqnarray}
iS^{-1}(p) &=& iS^{-1}_0(p) 
+ \sum_{n=0}^{N_{\rm KK}} \int \frac{d^4 q}{i(2\pi)^4} 
  \left[ -ig_n T^a \Gamma^M \right] S(q) 
  \left[ -ig_n T^a \Gamma^N \right] 
D_{MN}^{(n)} (q-p), \nonumber
\end{eqnarray}
where $S_0(p) = i/p\!\!\!/$, $S(p)$ and $D_{MN}^{(n)} (k)$ stand for 
a free fermion propagator, a full fermion propagator and a full $n$-th 
KK gauge boson propagator, respectively. We parameterize the full 
fermion propagator as $iS^{-1} (p) \equiv A(-p^2) p\!\!\!/ -B(-p^2)$ 
and solve the SD equation numerically in terms of $A$ and $B$. 

\begin{figure}[t]
\begin{center}
\begin{minipage}{0.46\linewidth}
\begin{center}
\epsfig{figure=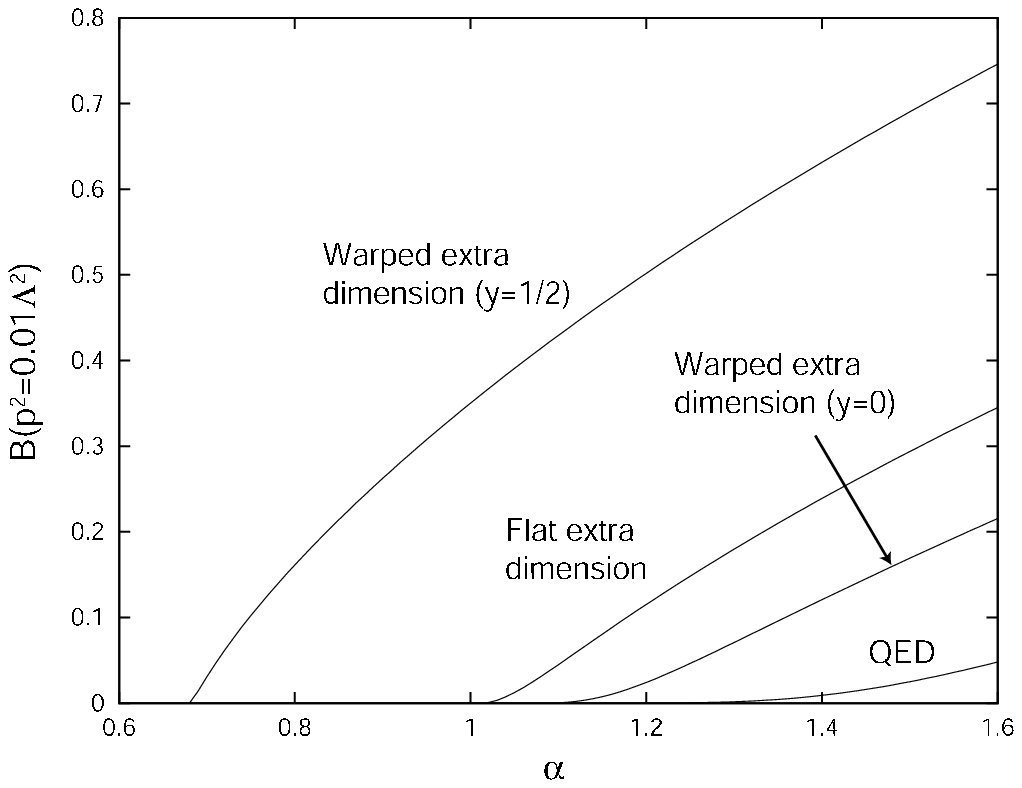,width=\linewidth} \\
Ladder SD ($N_{\rm KK}=1$)
\end{center}
\end{minipage}
\hfill
\begin{minipage}{0.46\linewidth}
\begin{center}
\epsfig{figure=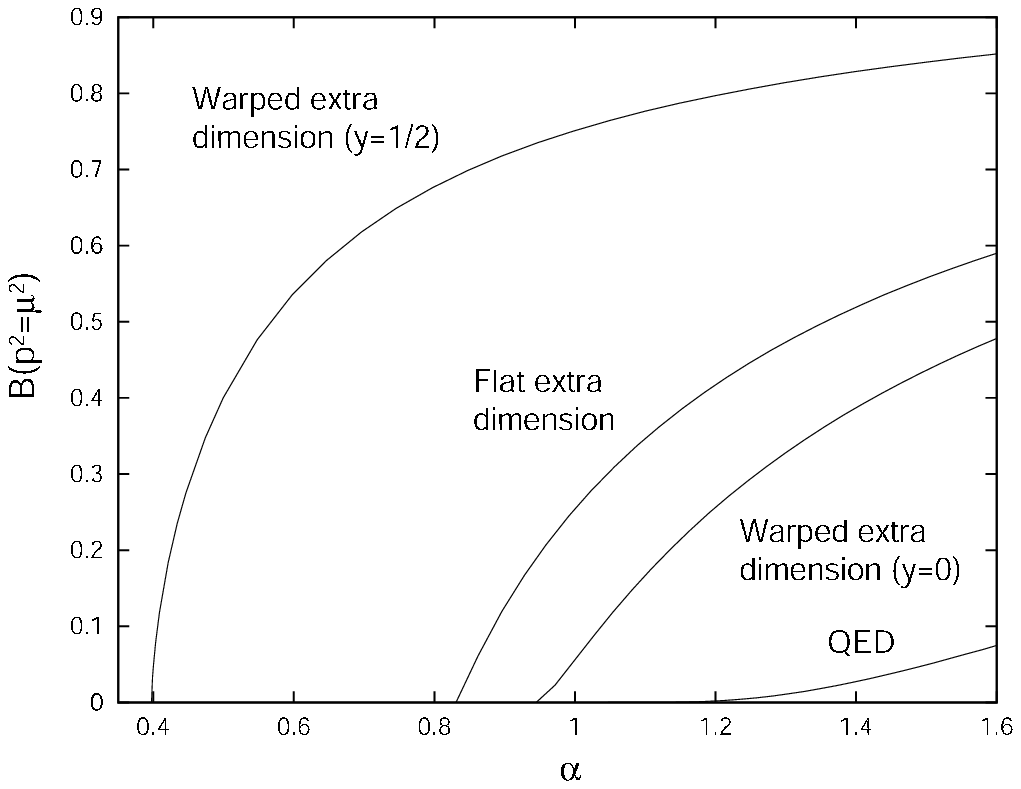,width=\linewidth} \\
Local four-fermion approximation ($N_{\rm KK}=1$)
\end{center}
\end{minipage}
\end{center}
\caption{Chiral phase structure on a boundary of bulk QED.}
\label{fig1}
\end{figure}

For orbifold QED (Abelian) theory with $M_c \lesssim \Lambda$, 
we apply so-called ladder approximation to the SD equation 
which utilizes a tree vertex and a tree gauge boson propagator. 
The results from the ladder SD and a local four-fermion approximation 
are shown in Fig.~\ref{fig1}. 
We conclude that the critical coupling $\alpha_c = g_c^2/4 \pi^2$ 
is less than the usual 4D one, which means the KK modes enhance DSB, and also 
claim that the local four-fermion approximation overestimates the KK effect 
on DSB, i.e., the KK propagation effect weakens DSB. 
The point (tree) vertex is sufficient? 
The $\alpha(\mu)$ runs indeed logarithmically, then it is 
negligible in QED (Abelian). And it runs in power low for $M_1<\mu<\Lambda$, 
then it is also negligible for $M_1 \sim \Lambda$ ($N_{\rm KK} \simeq 1$). 
However, for QCD (Yang-Mills) and/or $M_1 \ll \Lambda$ 
($N_{\rm KK} \gg 1$), we need to improve the ladder SD equation. 

\begin{figure}[t]
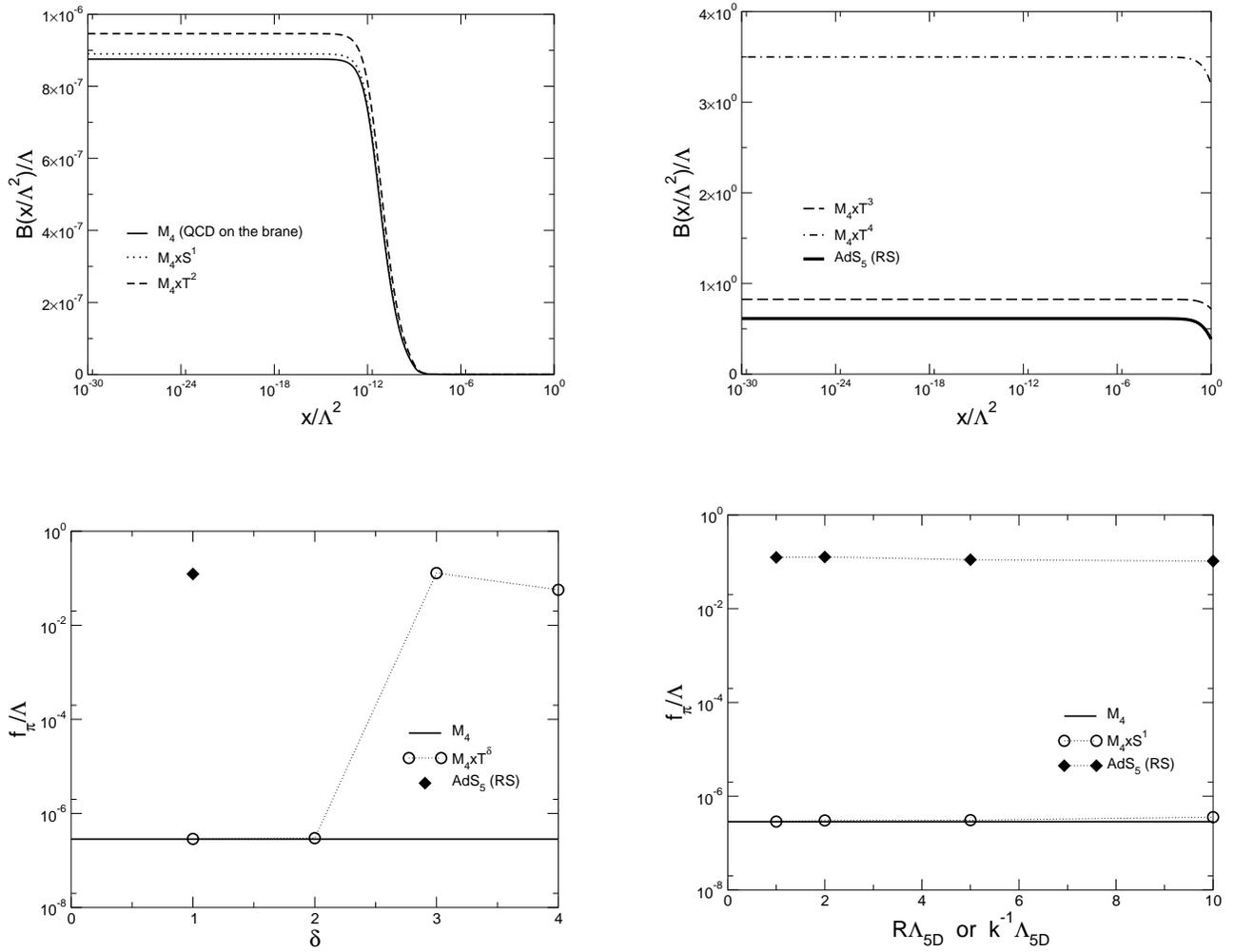

\begin{center}
\epsfig{figure=B_d0-d2.eps,width=0.46\linewidth} 
\hfill
\epsfig{figure=B_d3-rs.eps,width=0.46\linewidth}

\medskip

\medskip

\medskip

\medskip

\medskip

\epsfig{figure=Fpi.eps,width=0.46\linewidth} 
\hfill
\epsfig{figure=Fpi_R.eps,width=0.46\linewidth} 
\end{center}
\caption{Dynamical mass $B$ and $f_\pi$ on a boundary of bulk QCD 
($\Lambda_{\rm QCD}/\Lambda=2 \times 10^{-5}$).}
\label{fig2}
\end{figure}

Then, next we develop so-called improved ladder approximation 
(e.g., Ref.~\cite{Aoki:1990eq}) in the orbifold QCD.  
The one-loop perturbative running coupling is given by 
a truncated KK analysis~\cite{Dienes:1998vg} as 
\begin{eqnarray}
\frac{\pi}{4}\alpha^{-1} (z) &=& 
B \ln (z/\Lambda_{\rm QCD}^2)
-\widetilde{B} \left[ \ln (z/\mu_R^2) -\frac{2X_\delta}{\delta} 
\left\{ (z/\mu_R^2)^{\delta/2}-1 \right\} \right] \theta (z-\mu_R^2), 
\nonumber
\end{eqnarray}
where $\mu_R=1/R$, 
$B=\frac{1}{24C_2(F)}\left( \frac{11N_c-2N_{\rm f}}{3} \right) =9/16$ 
$(N_c=3,N_{\rm f}=3)$, $\widetilde{B} =\frac{1}{24C_2(F)} 
\left( \frac{11N_c-2\tilde{N}_{\rm f}}{3} \right) =11/16$ 
$(N_c=3,\tilde{N}_{\rm f}=0)$, 
$X_\delta = \frac{\pi^{\delta/2}}{\Gamma (1+\delta/2)}$ and 
$\delta$ is the number of extra dimensions.  
For a warped geometry we replace some quantities as 
$\Lambda = \Lambda_{\rm 5D} \to e^{-ky^\ast}\Lambda_{\rm 5D}$, 
$\mu_R \to \mu_{kR}=\pi e^{-\pi kR}k$, and 
$\sqrt{z} \to \sqrt{z}+\mu_{kR}/4$. 
For the orbifold QCD, the result of a fermion mass function $B(x)$ 
and the pion decay constant $f_\pi$ calculated from the $B(x)$ 
by using Pagels-Stocker formula are shown in Fig.~\ref{fig2}. 

From these figures we see that the behaviors of $B(x)$ 
are divided into two pieces. One is {\it Yang-Mills type} to which 
flat cases with $\delta=1,2$ and $y^\ast =0$ brane in warped case 
($w_0$) belong, where the zero mode gauge boson dominates DSB. 
Another is gauged {\it NJL type} to which flat cases with 
$\delta=3,4,\ldots$ and $y^\ast =\pi R$ brane in warped case ($w_\pi$) 
belong, where the gauge boson KK modes dominate DSB. 
The dynamical masses are given by $B(\Lambda_{\rm QCD}^2), 
f_\pi \sim \Lambda_{\rm QCD} \,(\delta=1,2;w_0)$, 
$\Lambda_{\rm 5D} \,(\delta=3,4)$, $e^{-\pi kR} \Lambda_{\rm 5D} \,(w_\pi)$. 
We observed almost no dependence on $N_{\rm KK}$ ($<10^2$), 
that means the power low running acts as a suppression factor for DSB. 

%\section{Summary}
In summary, a chiral phase structure (nonperturbative dynamics) 
is nontrivial on the orbifold. 
For the orbifold QED, $\alpha_c$ decreases as $N_{\rm KK} \sim R\Lambda$ 
increases. For the orbifold QCD, we have two sorts of the result. 
One is the Yang-Mills type ($\delta=1,2$ and $w_0$) and the other 
is the gauged NJL type ($\delta=3,4,\ldots$ and $w_\pi$). 
We may have to be careful about such dynamically generated masses 
in various kinds of phenomenological models on the orbifold. 
As future works, a detailed analysis with a nonlocal gauge fixing 
(e.g., Appendix in Ref.~\cite{Abe:2003hz}), with more precise 
one-loop running coupling (e.g., Ref.~\cite{Choi:2002ps}) and/or 
with different regularizations (e.g., Ref.~\cite{Abe:2002rj}) 
may be important.  It is also interesting to consider 
a quasi-localized fermion ($|M_{\rm kink}|<\Lambda$), 
a supersymmetric case (e.g., 5D ${\cal N}=1$ $\rightarrow$ 
4D ${\cal N}=2$, instanton, duality, etc.), and an application 
to a dynamical electroweak symmetry breaking 
(e.g., Ref.~\cite{Hashimoto:2000uk}). \\

This talk is based on Refs.~\cite{Abe:2003hz,Abe:2002yb,Abe:2001yi}. 

\pagebreak[0]


\begin{thebibliography}{99}

%\cite{Randall:1999ee}
\bibitem{Randall:1999ee}
L.~Randall and R.~Sundrum,
%``A large mass hierarchy from a small extra dimension,''
Phys.\ Rev.\ Lett.\  {\bf 83}, 3370 (1999)
[hep-ph/9905221].
%%CITATION = HEP-PH 9905221;%%

%\cite{Arkani-Hamed:1999dc}
\bibitem{Arkani-Hamed:1999dc}
N.~Arkani-Hamed and M.~Schmaltz,
%``Hierarchies without symmetries from extra dimensions,''
Phys.\ Rev.\ D {\bf 61}, 033005 (2000)
[hep-ph/9903417].
%%CITATION = HEP-PH 9903417;%%

%\cite{Antoniadis:1990ew}
\bibitem{Antoniadis:1990ew}
I.~Antoniadis,
%``A Possible New Dimension At A Few Tev,''
Phys.\ Lett.\ B {\bf 246}, 377 (1990).
%%CITATION = PHLTA,B246,377;%%

%\cite{Antoniadis:1993jp}
\bibitem{Antoniadis:1993jp}
I.~Antoniadis and K.~Benakli,
%``Limits on extra dimensions in orbifold compactifications of superstrings,''
Phys.\ Lett.\ B {\bf 326}, 69 (1994)
[arXiv:hep-th/9310151].
%%CITATION = HEP-TH 9310151;%%

%\cite{vonGersdorff:2001ak}
\bibitem{vonGersdorff:2001ak}
G.~von Gersdorff and M.~Quiros,
%``Supersymmetry breaking on orbifolds from Wilson lines,''
Phys.\ Rev.\ D {\bf 65}, 064016 (2002)
[hep-th/0110132].
%%CITATION = HEP-TH 0110132;%%

%\cite{Choi:2003di}
\bibitem{Choi:2003di}
K.~w.~Choi, D.~Y.~Kim, I.~W.~Kim and T.~Kobayashi,
%``Supersymmetry breaking in warped geometry,''
hep-ph/0305024.
%%CITATION = HEP-PH 0305024;%%

%\cite{Kawamura:2000ev}
\bibitem{Kawamura:2000ev}
Y.~Kawamura,
%``Triplet-doublet splitting, proton stability and extra dimension,''
Prog.\ Theor.\ Phys.\  {\bf 105}, 999 (2001)
[hep-ph/0012125].
%%CITATION = HEP-PH 0012125;%%

%\cite{Dobrescu:1998dg}
\bibitem{Dobrescu:1998dg}
B.~A.~Dobrescu,
%``Electroweak symmetry breaking as a consequence of compact dimensions,''
Phys.\ Lett.\ B {\bf 461}, 99 (1999)
[hep-ph/9812349].
%%CITATION = HEP-PH 9812349;%%

%\cite{Aoki:1990eq}
\bibitem{Aoki:1990eq}
K.~I.~Aoki, M.~Bando, T.~Kugo, M.~G.~Mitchard and H.~Nakatani,
%``Calculating The Decay Constant F(Pi),''
Prog.\ Theor.\ Phys.\  {\bf 84}, 683 (1990).
%%CITATION = PTPKA,84,683;%%

%\cite{Dienes:1998vg}
\bibitem{Dienes:1998vg}
K.~R.~Dienes, E.~Dudas and T.~Gherghetta,
%``Grand unification at intermediate mass scales through extra dimensions,''
Nucl.\ Phys.\ B {\bf 537}, 47 (1999)
[hep-ph/9806292].
%%CITATION = HEP-PH 9806292;%%

%\cite{Abe:2003hz}
\bibitem{Abe:2003hz}
H.~Abe,
%``Spontaneous and dynamical symmetry breaking in higher-dimensional space-time with boundary terms,''
hep-ph/0307004.
%%CITATION = HEP-PH 0307004;%%

%\cite{Choi:2002ps}
\bibitem{Choi:2002ps}
K.~w.~Choi and I.~W.~Kim,
%``One loop gauge couplings in AdS(5),''
Phys.\ Rev.\ D {\bf 67}, 045005 (2003)
[hep-th/0208071].
%%CITATION = HEP-TH 0208071;%%

%\cite{Abe:2002rj}
\bibitem{Abe:2002rj}
H.~Abe, T.~Kobayashi, N.~Maru and K.~Yoshioka,
%``Field localization in warped gauge theories,''
Phys.\ Rev.\ D {\bf 67}, 045019 (2003)
[hep-ph/0205344].
%%CITATION = HEP-PH 0205344;%%

%\cite{Hashimoto:2000uk}
\bibitem{Hashimoto:2000uk}
M.~Hashimoto, M.~Tanabashi and K.~Yamawaki,
%``Top mode standard model with extra dimensions,''
Phys.\ Rev.\ D {\bf 64}, 056003 (2001)
[hep-ph/0010260].
%%CITATION = HEP-PH 0010260;%%

%\cite{Abe:2002yb}
\bibitem{Abe:2002yb}
H.~Abe and T.~Inagaki,
%``Schwinger-Dyson analysis of dynamical symmetry breaking on a brane with bulk Yang-Mills theory,''
Phys.\ Rev.\ D {\bf 66}, 085001 (2002)
[hep-ph/0206282].
%%CITATION = HEP-PH 0206282;%%

%\cite{Abe:2001yi}
\bibitem{Abe:2001yi}
H.~Abe, K.~Fukazawa and T.~Inagaki,
%``Chiral phase transition of bulk Abelian gauge theories in the Randall-Sundrum brane world,''
Prog.\ Theor.\ Phys.\  {\bf 107}, 1047 (2002)
[hep-ph/0107125].
%%CITATION = HEP-PH 0107125;%%

\end{thebibliography}
\end{document}